\def\tipo{2}

\ifnum \tipo = 2
 \documentstyle[prl,twocolumn,aps,psfig]{revtex}
 \def\figsize{8cm}
 \def\figsiz1{7cm}
 \def \frontmatter{\twocolumn[\hsize\textwidth\columnwidth\hsize\csname@twocolumnfalse\endcsname}
 
\fi

\begin{document}
\draft
\frontmatter
\title{Magnitude and Sign Correlations in Heartbeat Fluctuations}
\author{Yosef~Ashkenazy$^{\text{a}}$, Plamen~Ch.~Ivanov$^{\text{a,b}}$, 
Shlomo~Havlin$^{\text{c}}$, Chung-K. Peng$^{\text{b}}$,\\
Ary L. Goldberger$^{\text{b}}$, and H. Eugene Stanley$^{\text{a}}$}
\address{ 
(a) Center for Polymer Studies and Department of Physics, Boston
University, Boston, Massachusetts 02215, USA\\
(b) Beth Israel Deaconess Medical Center,
Harvard Medical School, Boston, Massachusetts 02215, USA\\
(c) Dept. of Physics and Gonda Goldschmied Center, Bar-Ilan University, 
Ramat-Gan, Israel
}
\date{\today}
\maketitle
\begin{abstract}
{ We propose an approach for analyzing signals with long-range
correlations by decomposing the signal increment series into magnitude
and sign series and analyzing their scaling properties.  We show that
signals with identical long-range correlations can exhibit different
time organization for the magnitude and sign.  We find that the
magnitude series relates to the nonlinear properties of the original
time series, while the sign series relates to the linear properties.  We
apply our approach to the heartbeat interval series and find that the
magnitude series is long-range correlated, while the sign series is
anticorrelated and that both magnitude and sign series may 
have clinical applications.}
\end{abstract}
\ifnum \tipo = 2
]
\fi

\def\figureII{
\begin{figure}[thb]
\centerline{\psfig{figure=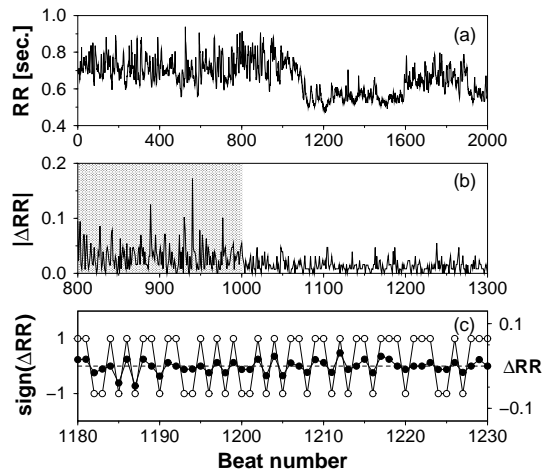,width=\figsiz1,angle=-90}}
{
\ifnum\tipo=2
\vspace*{0.0truecm}
\fi
\caption{\label{fig2} 
(a) An example of 2000 heartbeat (RR) intervals of a healthy subject
during daytime.   
(b) The magnitude series of a portion of the RR series (beat numbers
800-1300) shown in (a).
Patches of more ``volatile'' increments with large magnitude (beat 
numbers 800-1000) are followed by patches of less volatile increments 
with small magnitude (beat numbers 1000-1300), consistent with our 
quantitative conclusion that there is correlation in the magnitude time 
series.  
(c) The sign series ($\circ$), as well as the $\Delta RR$ series
($\bullet$) of a portion of the RR series (beat numbers 1180-1230)
shown in (a). The positive sign ($+1$) represents a positive
increment, while the negative sign ($-1$) represents a negative
increment in the RR series of interbeat intervals.  The tendency to
alternation between $+1$ and $-1$ is consistent
with our quantitative conclusion that there is (multiscale)
anticorrelation in the 
sign time series.
}}
\end{figure}
}

\def\figureIII{
\begin{figure}[thb]
\centerline{\psfig{figure=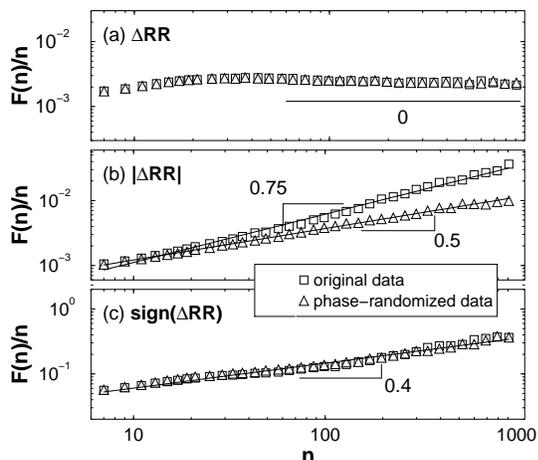,width=\figsiz1,angle=-90}}
{
\ifnum\tipo=2
\vspace*{-0.2cm}
\fi
\caption{\label{fig3} 
(a) Root mean square fluctuation, $F(n)$, for
$\approx 6$ hour record ($\approx 32,000 $ data points) for the
interbeat interval $RR_i$ series ($\Box$) of healthy subject
\protect\cite{remark3}. Here, $n$ 
indicates the time scale (in beat numbers) over which each measure is
calculated.  The scaling is obtained using 2nd order detrended
fluctuation analysis, and indicates long-range
anticorrelations in the heartbeat interval increment series $\Delta
RR_i$ {\protect\cite{Peng95}}. As expected, the scaling properties of
the heartbeat interval increment series remain unchanged after the
Fourier phase randomization ($\triangle$).
(b) The root mean square fluctuation of the integrated magnitude series
($\Box$) indicates long-range
correlations in the magnitude series $|\Delta RR_i|$ (group average
exponent of $\alpha-1 = 0.74\pm 0.08$ where 
$F(n)/n \propto n^{\alpha-1}$).  After
Fourier phase randomization of the interbeat interval increment series
we find random behavior with exponent 0.5
($\triangle$).  This change in the scaling (after removing the
nonlinear features in the time series) suggests that the magnitude
series carries information about the nonlinear properties of the
heartbeat dynamics.  
(c) The root mean square fluctuation of the integrated sign series
($\Box$) indicates anticorrelated behavior in ${\rm sign}(\Delta RR_i)$
(group average exponent of $\alpha-1 = 0.42\pm
0.03$ where $F(n)/n \propto n^{\alpha-1}$). 
The scaling properties of the sign series remain unchanged after the  
Fourier phase randomization ($\triangle$), which
suggests that the sign series relates to linear properties of
the heartbeat interval time series. 
We note the apparent crossovers at $n \approx 20$ beats and $n \approx
100$ beats.
A gradual loss of anticorrelation in the
sign series is observed at time scales larger than $n \approx 100$
beats. 
We note, however, that heartbeat increments derived from the original
time series are anticorrelated up to scales of thousands of heartbeats.
}}
\end{figure}
}

\def\figureIV{
\begin{figure}[thb]
\centerline{\psfig{figure=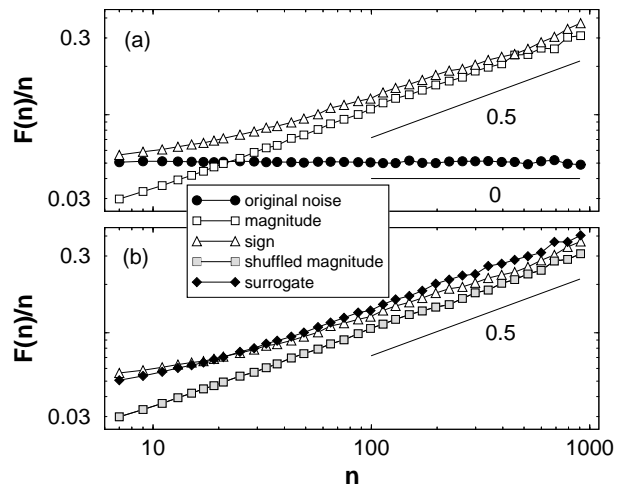,width=\figsize,angle=-90}}
\caption{\label{fig4}
(a) ($\bullet$) An example of anticorrelated noise (the scaling 
exponent of the increment series is 0) with ($\Box$) uncorrelated
magnitude series and ($\triangle$) uncorrelated sign series with
exponent 0.5 . (Note the
sign series is anticorrelated for $n<20$ and uncorrelated for $n>100$). 
(b) We shuffle the magnitude series from (a) (gray squares) and then
multiply its elements  
by the elements of the sign series from (a). The new surrogate series
(black diamonds) is uncorrelated. 
}
\end{figure}
}

\def\figureV{
\begin{figure}[thb]
\centerline{\psfig{figure=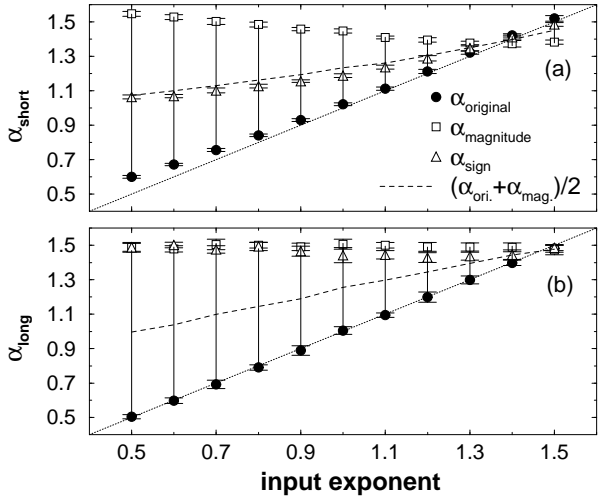,width=\figsize,angle=-90}}
\caption{\label{fig5}
(a) The relation between the scaling exponents of correlated noise, the
integrated magnitude series, and the integrated sign series for the
short range regime ($n<16$). We generate 10 series of length 32768 with
different correlations (input exponent) and
then calculate the scaling exponents of the original series ($\bullet$),
of the integrated magnitude series ($\Box$), and of the integrated sign
series ($\triangle$). In the figure we show the average $\pm$ 1 standard
deviation. The dashed line indicates the approximate empirical relation
between the different scaling exponents ($\alpha_{\rm sign} \approx
(\alpha_{\rm original}+ \alpha_{\rm magnitude})/2$). This empirical
approximation is good for the short range regime only ($n<16$).  
(b) Same as (a) for the long range regime ($n>64$). Here the
approximation does not hold anymore. In this regime the scaling
exponents of the magnitude and sign series are uncorrelated no matter
what is the exponent of the original series.}
\end{figure}
}

\def\tableI{
\begin{tabular}{|c||c|c|c|}
\hline
\multicolumn{4}{|c|}{magnitude}
	\\ \hline\hline
measure & healthy& heart failure & $p$ value \\ \hline
$\log_{10} F(n)$ & ${\bf -1.49}\pm{\bf 0.16}$ & ${\bf -1.92}\pm {\bf 0.17}$
& $1\times 10^{-7}$ \\ \hline
${\alpha}$ & ${\bf 1.74\pm 0.08}$ & ${\bf 1.66\pm 0.06}$ & 0.01 \\ \hline
$\alpha_1$ & $1.55\pm 0.08$ & $1.6\pm 0.08$ & 0.13 \\ \hline
$\alpha_2$ & $1.66\pm 0.08$ & $1.61\pm 0.08$ & 0.14 \\ \hline
$\alpha_3$ & ${\bf 1.82\pm 0.1}$ & ${\bf 1.71\pm 0.1}$ & $
4\times 10^{-3}$ \\ \hline
\hline
\multicolumn{4}{|c|}{sign}
	\\ \hline\hline
measure & healthy& heart failure& $p$ value \\ \hline
$\log_{10} F(n)$ & ${\bf 0.14}\pm {\bf 0.05}$ & ${\bf 0.02}\pm {\bf
0.06}$ & $1\times 10^{-6}$ \\ \hline
${\alpha}$ & ${1.42}\pm {0.03}$ & ${1.44}\pm {0.02}$ &
$0.08$ \\ \hline
$\alpha_1$ & ${\bf 1.43}\pm {\bf 0.12}$ & ${\bf 1.15}\pm {\bf 0.12}$ &
$7\times 10^{-7}$ \\ \hline
$\alpha_2$ & ${\bf 1.27}\pm {\bf 0.07}$ & ${\bf 1.41}\pm {\bf 0.07}$ &
$1\times 10^{-5}$ \\ \hline 
$\alpha_3$ & ${1.53\pm 0.065}$ & ${1.49\pm 0.04}$ & 0.04 \\ \hline 
\end{tabular}
\begin{table}
\caption{\label{table1}
Comparison of the statistics of the root mean square fluctuation,
$F(n)$ (calculated using the 2nd order detrended fluctuation analysis
method \protect\cite{Peng95} where $n$ is the time scale in beat
numbers over which each measure is calculated), and the scaling
exponents for 18 healthy subjects and 12 subjects  
with heart failure \protect\cite{MIT} (obtained from 6-hour records
during the day). 
The scaling features of the magnitude and sign change significantly
for the subjects with heart failure,
raising the possibility of bedside applications. ${\alpha}$ is
the best fit to the range $6<n<1024$.
$F(n)$ is estimated at the crossover position ($n=16$)
(Fig. \ref{fig3}b) where the largest separation between the two groups
is estimated. Since we observe two apparent crossovers in the 
scaling behavior of the sign series, we calculate the
scaling exponents in three different 
regions : (i) the short range regime for time scales $6<n<16$ with
scaling exponent, $\alpha_1$, (ii) the intermediate regime for time scales
$16\le n\le 64$ with scaling exponent, $\alpha_2$, (iii) and the long
range regime for time scales $64< n\le 1024$ with scaling exponent,
$\alpha_3$. For each measure, the group average $\pm$ 1 standard
deviation is presented. The values which show highly significant differences
($p \le 0.01$ by Student's $t$-test) between the healthy and
heart failure groups are indicated in boldface. 
We note, surprisingly, that the short range and the intermediate range
scaling exponents $\alpha_1$ and $\alpha_2$ of the sign series 
may provide even more robust separation between healthy and heart
failure compared to previous reports \protect\cite{Peng95} based on the
scaling exponents of the original heartbeat series.
}
\end{table}
}

A broad class of physical and biological systems exhibits
complex dynamics, associated with the presence of many
components interacting over a wide range of time or
space scales. These often-competing interactions may
generate an output signal with fluctuations that appear ``noisy'' and
``erratic'' but reveal scale-invariant structure. 
One general approach to study these systems is to 
analyze the ways that such fluctuations obey scaling laws
\cite{Shlesinger,Takayasu,Kobayashi}. 

Here, we take into account that the fluctuations in the dynamical output
of any system can be characterized by their magnitude (absolute value)
and their direction (sign). These two quantities reflect the underlying
interactions in a system --- the resulting ``force'' of these
interactions at each moment determines the magnitude and the direction
of the fluctuations. For an important representative of complex dynamics
--- human heartbeat intervals --- we find unexpected results for the
time ordering of the heartbeat interval fluctuations by studying the
scaling properties of their magnitude and sign.  We also demonstrate
that fluctuations following identical long-range correlations can
exhibit very different time ordering for the magnitude and sign.

We consider the time series formed by consecutive cardiac interbeat
intervals (Fig. \ref{fig2}a) and focus on the correlations in the time
{\it increments} between consecutive beats.  This time series is of
general interest, in part because it is the output of a complex
integrated control system, including competing stimuli from the
neuroautonomic nervous system \cite{Berne}.  These stimuli modulate the
rhythmicity of the heart's intrinsic pacemaker, leading to complex
fluctuations.  Previous reports indicate that these fluctuations exhibit
scale-invariant properties, and are anticorrelated over a broad range of
time scales (i.e., the power spectrum follows a power-law where the
amplitudes of the higher frequencies are dominant)
\cite{remark1,Peng95}.

The time series of the fluctuations in heartbeat intervals can be
``decomposed'' into two different time series.  We analyze separately
the time series formed by the magnitude and the sign of the increments
in the time intervals between successive heartbeats
(Fig. \ref{fig2}b,c).  We use 2nd order detrended fluctuation analysis
\cite{Peng95} (and not the conventional power spectrum) since it has the
ability to accurately estimate correlations in the heartbeat
fluctuations even when they are masked by linear trends \cite{remark2a}.
We find for 
each subject in a group of 18 healthy individuals \cite{MIT}, that the
time series of the magnitudes exhibits correlated behavior
(Fig. \ref{fig3}b) (unlike the original 
\ifnum\tipo=2
  \figureII
  \figureIII
\fi
heartbeat increment time series,
which is anticorrelated, Fig. \ref{fig3}a). 
The sign series, however, exhibits 
anticorrelated behavior (Fig. \ref{fig3}c) \cite{remark3}.  Correlation
in the magnitude series indicates that an increment with large magnitude
is more likely to be followed by an increment with large magnitude.
Anticorrelation in the sign series indicates that a positive increment
is more likely to be followed by a negative increment. Our result for
the temporal organization of heartbeat fluctuations thus suggests that,
under healthy conditions, a large increment in the positive direction is
more likely to be followed by a large increment in the negative
direction.  We find that this empirical ``rule'' holds over a broad
range of time scales from several up to hundreds of beats
(Fig. \ref{fig3})
\cite{remark2}. 

To show that fluctuations following an identical scaling law can
exhibit different time ordering for the magnitude and sign,
we perform a Fourier transform on a heartbeat interval increment time
series, preserving the amplitudes of the Fourier transform but
randomizing the Fourier phases. Then we perform an inverse Fourier
transform to create a surrogate series. This procedure eliminates
non-linearities, preserving only the linear features (i.e. two-point
correlations) of the original time series \cite{Panter}.  The new
surrogate series has the {\it same} power spectrum as the original
heartbeat interval increment time series, with a scaling exponent
indicating long-range anticorrelations in the interbeat increments
(Fig. \ref{fig3}a). Our
analysis of the sign time series derived from this surrogate signal
shows scaling behavior almost identical to the one for the sign series
from the original data (Fig. \ref{fig3}c). However, the magnitude time
series derived from the surrogate (linearized) signal exhibits {\it
uncorrelated} behavior --- a significant change from the strongly {\it
correlated} behavior observed for the original magnitude series
(Fig. \ref{fig3}b).  Thus, the increments in the surrogate series do not
follow the empirical ``rule'' observed for the original heartbeat
series, although these increments follow a scaling law identical to the
original heartbeat increment series.  
Moreover, our results raise the
interesting possibility that the magnitude series carries information
about the nonlinear properties of the heartbeat series, while the sign
series relates importantly to linear properties.

Next, we investigate the relation between the scaling exponent of the
original series and the scaling exponents 
\ifnum\tipo=2
  \figureIV
  \figureV
\fi
of the magnitude and the sign
series. For this purpose, 
we test our approach on well-defined signals 
with built-in correlated behavior 
that show uncorrelated behavior for the magnitude and sign. First,
we consider a particular example of correlated noise with scaling
exponent equal to 1, for which the increment series is anticorrelated with
scaling exponent equal to 0 (Fig. \ref{fig4}a). 
Surprisingly, at large time scales, we find that the magnitude series and
the sign series of the increments exhibit uncorrelated behavior (scaling
exponent of 0.5) although the original increment series, which is the
multiplication of the elements of these two series, is strongly
anticorrelated.
Moreover, we find that for linear colored noise with correlation
exponent less than 1.5 (i.e., with anticorrelations for the increment
series), the magnitude and sign series of the increments 
are uncorrelated (Fig. \ref{fig5}b). Next, we shuffle the magnitude
series by randomly exchanging pairs of elements. After multiplication of
the elements of the shuffled magnitude 
series with the elements of the sign series, we find that the resulting
time series is uncorrelated, in contrast to the original increments time
series which is strongly anticorrelated. Note that the scaling exponents of
the magnitude and sign series remain the same as before the shuffling
(Fig. \ref{fig4}b). This test indicates that the correlations in a time
series are not related to the correlations in the magnitude and sign
series, but rather to the particular pairing of the elements of the
magnitude and sign series. 

At small time scales, however, we find an empirical
approximate relation for the scaling exponents (Fig. \ref{fig5}a), 
$
\alpha_{\rm sign} \approx {1 \over 2}(\alpha_{\rm original}+\alpha_{\rm
magnitude}). 
$
We observe that for the heartbeat series this relation is valid
over a larger range of scales (i.e., for time scales $n<100$).

\ifnum\tipo=2
 \tableI
\fi
Finally, we test our analysis on a group of 12 subjects with
congestive heart failure \cite{MIT}. Compared to the healthy subjects,
the magnitude exhibits weaker correlations with a scaling exponent
closer to the exponent of an uncorrelated series. The change in the
magnitude exponent for the heart failure subjects is consistent with a
previously reported loss of nonlinearity with disease
\cite{Kurths95,Sugihara}. 
The sign time series of heart failure subjects shows scaling behavior
similar to the one 
observed in the original time series, but significantly different from
the healthy subjects (Table \ref{table1}).  

We conclude that series with identical correlation properties
can have completely different time ordering which can be
characterized by different
scaling exponents for the magnitude
and sign series. Moreover, we show that the magnitude series carries
information regarding the nonlinear properties of the original series
while the sign series carries information regarding the linear
properties of the original series. 
The significant decrease in the short-range scaling exponent for
the sign series in heart failure may be related to perturbed vagal control
affecting relatively high frequency fluctuations. The decrease of the
long-range scaling exponent for the magnitude series of the heart failure
patients indicates weaker correlations and loss of nonlinearity which may
be related to impaired feedback mechanisms of neurohormonal cardiac
regulation. 
Because information obtained by decomposing the
original heartbeat time series into magnitude and sign time series
likely reflects aspects of neuroautonomic regulation, this type of
analysis may help address the challenge of developing realistic models
of heart rate control in health and disease.

Partial support was provided by the NIH/National Center for Research
Resources (P41 RR13622), the 
Mathers Charitable Foundation, the National Aeronautics and Space
Administration (NASA), the Centers for Disease Control, the Fetzer
Institute, and the Israel-USA Binational Science Foundation. 
We thank two referees for their helpful comments.

\ifnum\tipo=2
\vspace*{-0.6cm}
\fi

\ifnum\tipo=1
  \figureII
  \figureIII
  \figureIV
  \figureV
  \tableI
\fi


\begin{references}
\bibitem{remark1} By long-range anticorrelations we also mean that the
root mean square fluctuations function of the integrated series is 
proportional to $n^\alpha$ where $n$ is the window scale and the scaling
exponent $\alpha$ is smaller than 0.5. In contrast, for uncorrelated
behavior $\alpha=0.5$, while for correlated behavior $\alpha>0.5$. 
In the present study we integrate the series before applying the scaling
analysis and thus $\alpha=1.5$ indicates uncorrelated behavior while
$\alpha>1.5$ ($\alpha<1.5$) indicates correlated (anticorrelated)
behavior.
\bibitem{remark3} The magnitude/sign decomposition consists of the
following steps: (i) given $RR_i$ series we create the increment series,
$\Delta RR_i=RR_{i+1}-RR_i$; (ii) we decompose the increment series
into a magnitude 
series ($|\Delta RR|$) and a sign series (${\rm sign}(\Delta RR)$); (iii)
to avoid artificial trends we subtract from the magnitude and sign
series their average; (iv) because of limitations in the accuracy of the
detrended fluctuation analysis method for estimating the scaling
exponents of anticorrelated signals ($\alpha <0.5$), 
we integrate the magnitude and sign series;
(v) we perform a scaling analysis using 2nd order detrended
fluctuation analysis on the integrated magnitude and sign series;
(vi) to obtain the scaling exponents for the magnitude and sign series we
measure the slope of $F(n)/n$ on a log-log plot, where $F(n)$ is the
root mean square fluctuation function and $n$ is the scale of analysis
(in beat numbers).
\end{references}
\end{document}